\documentclass[a4paper]{spie}
\usepackage{amsfonts}
\usepackage[T1]{fontenc}
\usepackage[mathcal]{eucal}
\usepackage{graphicx}
\usepackage{texdraw}
\usepackage[pdfborder={0 0 0}]{hyperref}

\def\abs#1{\left|#1\right|}
\def\bkt#1{\left(#1\right)}
\def\od#1{\!\bkt{#1}}
\def\bref#1{(\ref{#1})}
\def\smatrix#1{\left[\matrix{#1}\right]}
\def\lid#1#2#3{\int_{#2}^{#3}\! d{#1}\,}
\def\real#1{\mathrm{Re}\left\{#1\right\}}

\def\pol{\epsilon}
\def\NOS#1{N\od{#1}}
\def\DOSsign{dN}
\def\DOS#1{\DOSsign\od{#1}}
\def\DOSfs#1{\DOSsign_{fs}\od{#1}}
\def\mslett{\rho}
\def\ms#1#2{\mslett_{#1}\od{#2}}
\def\Qe#1#2{\Lambda_{#1}\od{#2}}

\title{Simple model of the density of states in 1D photonic crystal}
\author{Adam Rudzi\'nski, Anna Tyszka-Zawadzka, Pawe\l{} Szczepa\'nski \skiplinehalf Institute of Microelectronics and Optoelectronics, Warsaw University of Technology, Warsaw, Poland}

\begin{document}
\bibliographystyle{spiebib}

\maketitle

\abstract
In this paper, we present a simple, yet versatile, analytical model of
one-dimensional photonic crystal (1D~PC). In our
theoretical model, we take into account direction of propagation and
therefore do not neglect anisotropic nature of photonic crystals.
We derive analytical expressions for mode spectrum and density
of states in 1D photonic crystal. With those formulas, we obtain mode
spectrum characteristics, which depict formation of photonic band gap
and reveal properties of photonic crystals.

\section{Introduction}
Photonic crystals (PCs), proposed by Yablonovitch\cite{Yablon}
and John\cite{John}, represent a novel class of optical materials
which allow to control the flow of electromagnetic radiation or to
modify light-matter interaction. These artificial structures are
characterized by one-, two- or three-dimensional arrangements of
dielectric material which lead to the formation of an energy band
structure for electromagnetic waves propagating in them. One of
the most interesting features of photonic crystals is associated with
the fact that PCs may exhibit frequency ranges over which ordinary
linear propagation is forbidden, irrespective of direction. These
photonic band gaps (PBGs) lend  themselves to numerous diversified
applications (in linear, nonlinear and quantum optics). For
instance, PBG structures with line defects can be used for guiding
light. Similarly, as it has been predicted and confirmed
experimentally, photonic crystals allow to modify spontaneous
emission rate due to the modification of density of quantum
states. In particular, it is well known that the density of states
grows at the edge of the photonic band gap of the PC. This allows
us to predict a higher optical gain, but on the other hand a higher level
of noise in light generated in PC-lasers operated at a frequency near
the band gap.

It is our purpose here to present a simple, yet versatile,
analytical model of one-dimensional photonic crystal.
The investigated photonic crystal consists of $N$
elementary cells with alternating layers of different indices of
refraction and different widths. Presented model takes into
consideration direction of propagation and therefore does not
neglect anisotropic nature of photonic crystals. We derive
analytical expressions for mode spectrum and density of states
that allow to analyze properties of one-dimensional photonic
crystals, especially inhibition or amplification of spontaneous
emission due to modification of density of states. The described
model allows to examine basic physics behind PBG structures and as
an analytical model - can provide an excellent starting point in
numerical calculations for much more complicated structures.

\section{Effective resonance approach to photonic crystals}
\begin{figure}[!htb]
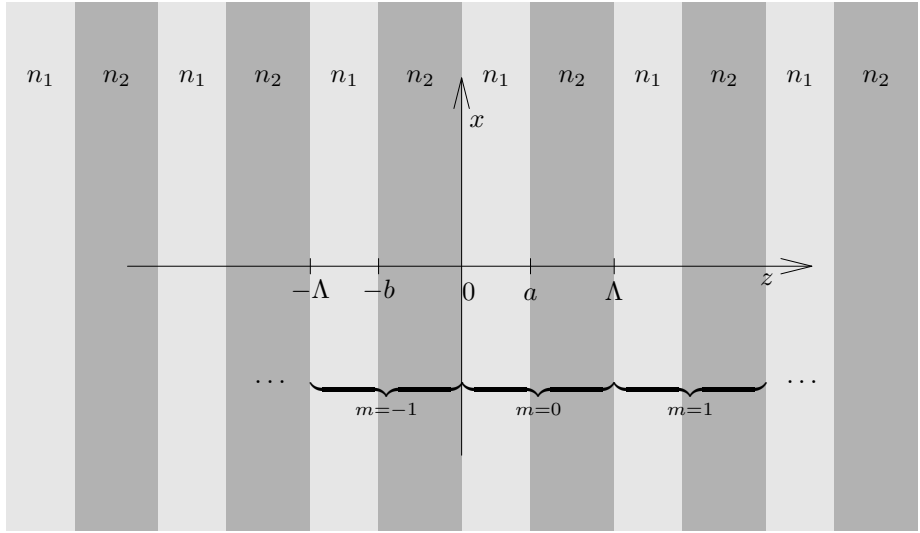

\centertexdraw
{
    \drawdim cm \linewd .01 \textref h:C v:C \arrowheadtype t:V
    \rlvec(0 7) \rlvec(0.9 0) \rlvec(0 -7) \rlvec(-0.9 0) \ifill f:0.9 \rmove(.45 6) \htext{$n_1$} \rmove(.45 -6)
    \rlvec(0 7) \rlvec(1.1 0) \rlvec(0 -7) \rlvec(-1.1 0) \ifill f:0.7 \rmove(.55 6) \htext{$n_2$} \rmove(.55 -6)
    \rlvec(0 7) \rlvec(0.9 0) \rlvec(0 -7) \rlvec(-0.9 0) \ifill f:0.9 \rmove(.45 6) \htext{$n_1$} \rmove(.45 -6)
    \rlvec(0 7) \rlvec(1.1 0) \rlvec(0 -7) \rlvec(-1.1 0) \ifill f:0.7 \rmove(.55 6) \htext{$n_2$} \rmove(.55 -6)
    \rlvec(0 7) \rlvec(0.9 0) \rlvec(0 -7) \rlvec(-0.9 0) \ifill f:0.9 \rmove(.45 6) \htext{$n_1$} \rmove(.45 -6)
    \rlvec(0 7) \rlvec(1.1 0) \rlvec(0 -7) \rlvec(-1.1 0) \ifill f:0.7 \rmove(.55 6) \htext{$n_2$} \rmove(.55 -6)
    \rlvec(0 7) \rlvec(0.9 0) \rlvec(0 -7) \rlvec(-0.9 0) \ifill f:0.9 \rmove(.45 6) \htext{$n_1$} \rmove(.45 -6)
    \rlvec(0 7) \rlvec(1.1 0) \rlvec(0 -7) \rlvec(-1.1 0) \ifill f:0.7 \rmove(.55 6) \htext{$n_2$} \rmove(.55 -6)
    \rlvec(0 7) \rlvec(0.9 0) \rlvec(0 -7) \rlvec(-0.9 0) \ifill f:0.9 \rmove(.45 6) \htext{$n_1$} \rmove(.45 -6)
    \rlvec(0 7) \rlvec(1.1 0) \rlvec(0 -7) \rlvec(-1.1 0) \ifill f:0.7 \rmove(.55 6) \htext{$n_2$} \rmove(.55 -6)
    \rlvec(0 7) \rlvec(0.9 0) \rlvec(0 -7) \rlvec(-0.9 0) \ifill f:0.9 \rmove(.45 6) \htext{$n_1$} \rmove(.45 -6)
    \rlvec(0 7) \rlvec(1.1 0) \rlvec(0 -7) \rlvec(-1.1 0) \ifill f:0.7 \rmove(.55 6) \htext{$n_2$} \rmove(-5.45 -5)
    \ravec(0 5) \rmove(.1 -.5) \textref h:L v:T \htext{$x$} \rmove(-1 .5) \rmove(-3.5 -2.5)
    \ravec(9 0) \rmove(-.5 -.1) \textref h:R v:T \htext{$z$}
    \textref h:C v:B \rmove(-6.1 .2) \rlvec(0 -.2) \rmove(0 -.35) \htext{$-\Lambda$} \rmove(.9 .55)
    \rlvec(0 -.2) \rmove(0 -.35) \htext{$-b$} \rmove(1.1 0)
    \textref h:L v:B \htext{$0$} \textref h:C v:B \rmove(.9 .55)
    \rlvec(0 -.2) \rmove(0 -.35) \htext{$a$} \rmove(1.1 .55)
    \rlvec(0 -.2) \rmove(0 -.35) \htext{$\Lambda$} \rmove(.9 .55)

    \move(3.5 2) \textref h:C v:T \htext{\ldots}
    \rmove(1.5 0) \htext{$\underbrace{\rule{2cm}{0pt}}_{m=-1}$}
    \rmove(2 0) \htext{$\underbrace{\rule{2cm}{0pt}}_{m=0}$}
    \rmove(2 0) \htext{$\underbrace{\rule{2cm}{0pt}}_{m=1}$}
    \rmove(1.5 0) \htext{\dots}
}
\caption{Structure of one-dimensional photonic crystal (section).}
\label{pc1d}
\end{figure}

The simplest case of a photonic crystal is one-dimensional photonic crystal. It is a structure
built of alternating layers of linear, uniform and isotropic materials of two different refractive indices $n_1$ and $n_2$.
Its section in $xz$ plane is shown on figure \ref{pc1d}. The structure has rotational symmetry around $z$ axis,
all layers extend towards infinity in $x$ and $y$ directions (direction $y$ is perpendicular to section
shown on figure \ref{pc1d}). We will consider a~photonic crystal built of finite number of layers.
A primitive cell in this case is a pair of subsequent layers with refractive indices $n_1$ and $n_2$.
The number of primitive cells ``right'' from $z=0$ plane will be $N_R$ and to the ``left'' will be $N_L$,
where directions ``right'' and ``left'' are in accordance with those on figure \ref{pc1d}.
We will introduce primitive cell index $m$ as depicted on the figure: layers with $z\in\smatrix{0,\Lambda}$
constitute primitive cell with $m=0$, and the next (along with the $z$ axis) pair has $m=1$, the
first one has $m=N_R-1$, the first primitive cell ``left'' from $z=0$ plane has $m=-1$ and
the index of the last one is $m=-N_L$. We also assume (for simplicity), that the crystal
is surrounded by a material with refractive index $n_1$.

\begin{figure}[!htb]
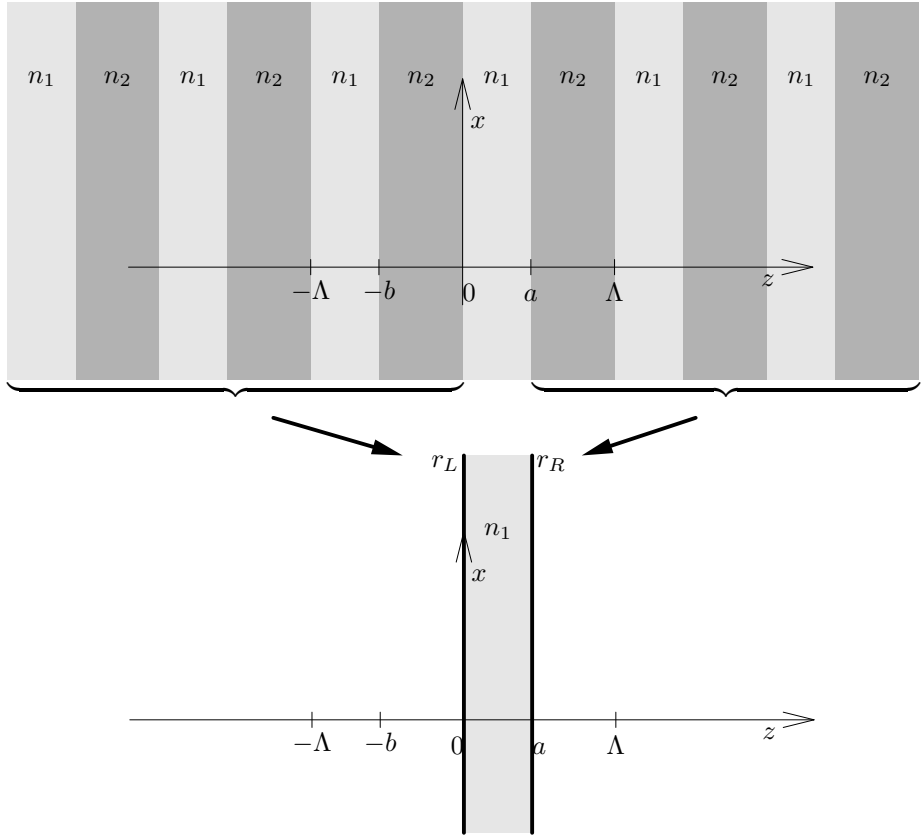

\centertexdraw
{
    \drawdim cm \linewd .01 \textref h:C v:C \arrowheadtype t:V

    \move(6 0)
    \rlvec(0 5) \rlvec(0.9 0) \rlvec(0 -5) \rlvec(-0.9 0) \ifill f:0.9 \rmove(.45 4) \htext{$n_1$}

    \move(0 0) \rmove(6 1)
    \ravec(0 3) \rmove(.1 -.5) \textref h:L v:T \htext{$x$} \rmove(-1 .5) \rmove(-3.5 -2.5)
    \ravec(9 0) \rmove(-.5 -.1) \textref h:R v:T \htext{$z$} \rmove(.5 .1)
    \textref h:C v:B \rmove(-6.6 .1) \rlvec(0 -.2) \rmove(0 -.35) \htext{$-\Lambda$} \rmove(.9 .55)
    \rlvec(0 -.2) \rmove(0 -.35) \htext{$-b$} \rmove(1.1 0)
    \textref h:R v:B \htext{$0$} \rmove(.9 .55)
    \rlvec(0 -.2) \rmove(0 -.35) \textref h:L v:B \htext{$a$} \textref h:C v:B \rmove(1.1 .55)
    \rlvec(0 -.2) \rmove(0 -.35) \htext{$\Lambda$} \rmove(.9 .55)

    \move(0 6) \textref h:C v:C
    \rlvec(0 5) \rlvec(0.9 0) \rlvec(0 -5) \rlvec(-0.9 0) \ifill f:0.9 \rmove(.45 4) \htext{$n_1$} \rmove(.45 -4)
    \rlvec(0 5) \rlvec(1.1 0) \rlvec(0 -5) \rlvec(-1.1 0) \ifill f:0.7 \rmove(.55 4) \htext{$n_2$} \rmove(.55 -4)
    \rlvec(0 5) \rlvec(0.9 0) \rlvec(0 -5) \rlvec(-0.9 0) \ifill f:0.9 \rmove(.45 4) \htext{$n_1$} \rmove(.45 -4)
    \rlvec(0 5) \rlvec(1.1 0) \rlvec(0 -5) \rlvec(-1.1 0) \ifill f:0.7 \rmove(.55 4) \htext{$n_2$} \rmove(.55 -4)
    \rlvec(0 5) \rlvec(0.9 0) \rlvec(0 -5) \rlvec(-0.9 0) \ifill f:0.9 \rmove(.45 4) \htext{$n_1$} \rmove(.45 -4)
    \rlvec(0 5) \rlvec(1.1 0) \rlvec(0 -5) \rlvec(-1.1 0) \ifill f:0.7 \rmove(.55 4) \htext{$n_2$} \rmove(.55 -4)
    \rlvec(0 5) \rlvec(0.9 0) \rlvec(0 -5) \rlvec(-0.9 0) \ifill f:0.9 \rmove(.45 4) \htext{$n_1$} \rmove(.45 -4)
    \rlvec(0 5) \rlvec(1.1 0) \rlvec(0 -5) \rlvec(-1.1 0) \ifill f:0.7 \rmove(.55 4) \htext{$n_2$} \rmove(.55 -4)
    \rlvec(0 5) \rlvec(0.9 0) \rlvec(0 -5) \rlvec(-0.9 0) \ifill f:0.9 \rmove(.45 4) \htext{$n_1$} \rmove(.45 -4)
    \rlvec(0 5) \rlvec(1.1 0) \rlvec(0 -5) \rlvec(-1.1 0) \ifill f:0.7 \rmove(.55 4) \htext{$n_2$} \rmove(.55 -4)
    \rlvec(0 5) \rlvec(0.9 0) \rlvec(0 -5) \rlvec(-0.9 0) \ifill f:0.9 \rmove(.45 4) \htext{$n_1$} \rmove(.45 -4)
    \rlvec(0 5) \rlvec(1.1 0) \rlvec(0 -5) \rlvec(-1.1 0) \ifill f:0.7 \rmove(.55 4) \htext{$n_2$} \rmove(-5.45 -3)
    \ravec(0 3) \rmove(.1 -.5) \textref h:L v:T \htext{$x$} \rmove(-1 .5) \rmove(-3.5 -2.5)
    \ravec(9 0) \rmove(-.5 -.1) \textref h:R v:T \htext{$z$} \rmove(.5 .1)
    \textref h:C v:B \rmove(-6.6 .1) \rlvec(0 -.2) \rmove(0 -.35) \htext{$-\Lambda$} \rmove(.9 .55)
    \rlvec(0 -.2) \rmove(0 -.35) \htext{$-b$} \rmove(1.1 0)
    \textref h:L v:B \htext{$0$} \textref h:C v:B \rmove(.9 .55)
    \rlvec(0 -.2) \rmove(0 -.35) \htext{$a$} \rmove(1.1 .55)
    \rlvec(0 -.2) \rmove(0 -.35) \htext{$\Lambda$} \rmove(.9 .55)

    \linewd 0.05 \move(6 0) \rlvec(0 5) \textref h:R v:T \htext{$r_L\,$} \rmove(0.9 0) \textref h:L v:T \htext{$\,r_R$} \rlvec(0 -5)

    \move(0 6) \htext{$\underbrace{\rule{6cm}{0pt}}$} \rmove(6.9 0) \htext{$\underbrace{\rule{5.1cm}{0pt}}$}
    \arrowheadtype t:F \rmove(-3.9 0) \rmove(.5 -.5) \ravec(1.7 -.5) \rmove(-2.2 1) \rmove(6.45 0) \rmove(-.4 -.5) \ravec(-1.5 -.5)
}
\caption{A layer of one-dimensional photonic crystal as a resonator.}
\label{pc1dres}
\end{figure}

The idea of the described method\cite{mgr,inz} is presented on figure \ref{pc1dres}. It aims to determine
effective reflection from a part of photonic crystal on a layer's boundary and ``replace'' it
with a mirror of the same reflection. Thus, we can say, that chosen crystal's layer
in which the density of states is investigated, is inside an effective FP cavity.
This method can describe properties of finite photonic
crystals, and allows to examine changes of these properties in different layers of the crystal.
In general case of photonic crystal, precise description of density of states without approximations requires
difficult and time consuming calculations. However, for the simplest case of
one-dimensional photonic crystal it is possible to follow the derivation given
below in the further part of the paper.

In our model we assume that in each of layers of the photonic crystal Maxwell equations\cite{Jack} have solutions in form of plane-waves:
\begin{equation}\vec E=\vec E_0e^{i\vec k\vec r-i\omega t}\end{equation}
with wave vector $\vec k$ bound with angular frequency $\omega$ by dispersion relation
\begin{equation}\vec k^2=\frac{n^2\omega^2}{c^2}\end{equation}
where $c$ is speed of light in vacuum and $n$ is refractive index of the material of the layer.
We can relate fields in different layers using continuity conditions, what is especially simple
in case of one-dimensional photonic crystal\cite{Yeh}. In every $n_1$ layer we represent electric field as a superposition
of coupled plane-waves:
\begin{eqnarray}E_{m,1,\pol}=a_{m,\pol}e^{ik_xx+i\beta_1z-i\omega t}+b_{m,\pol}e^{ik_xx-i\beta_1z-i\omega t} & , & z\in\smatrix{m\Lambda,m\Lambda+a}\end{eqnarray}
and similarly in every $n_2$ layer:
\begin{eqnarray}E_{m,2,\pol}=c_{m,\pol}e^{ik_xx+i\beta_2z-i\omega t}+d_{m,\pol}e^{ik_xx-i\beta_2z-i\omega t} & , & z\in\smatrix{m\Lambda+a,\bkt{m+1}\Lambda}\end{eqnarray}
where $\pol$ denotes polarization $TE$ (electric field perpendicular to plane of incidence) or $TM$
(electric field in plane of incidence) and
\begin{eqnarray}\beta_j=\sqrt{\frac{n_j^2\omega^2}{c^2}-k_x^2}& , & j=1,2\label{beta}\end{eqnarray}
Continuity conditions lead to equation relating amplitudes in consecutive primitive cells:
\begin{equation}\smatrix{a_{m+1,\pol} \cr b_{m+1,\pol}}=M_{m,\pol}\smatrix{a_{m,\pol} \cr b_{m,\pol}}\end{equation}
where $M_{m,\pol}$ is \emph{translation matrix} of $m$th primitive cell. It is an obvious conclusion, that if
amplitudes of plane-waves outside the photonic crystal are indexed with $m=N_R$ for ``right'' and
$m=-N_L-1$ for ``left'', then
\begin{eqnarray}& & \smatrix{a_{N_R,\pol} \cr b_{N_R,\pol}}=\bkt{M_{N_R-1,\pol}M_{N_R-2,\pol}\ldots M_{1,\pol}M_{0,\pol}}\smatrix{a_{0,\pol} \cr b_{0,\pol}}\label{Mgenright}\\
& & \smatrix{a_{-N_L-1,\pol} \cr b_{-N_L-1,\pol}}=\bkt{M_{-N_L,\pol}M_{-N_L+1,\pol}\ldots M_{-2,\pol}M_{-1,\pol}}^{-1}\smatrix{a_{0,\pol} \cr b_{0,\pol}}\label{Mgenleft}\end{eqnarray}
with the amplitudes outside the crystal bound by external conditions.
\bref{Mgenright} and \bref{Mgenleft} are general relations, that allow to perform calculations for
defected structures. For a perfect photonic crystal we have
\begin{eqnarray}M_{m,\pol}=M_{m',\pol}\equiv M_\pol & , & m,m'\in\smatrix{-N_L,N_R-1}\end{eqnarray}
with
\begin{equation}M_\pol=\smatrix{A_\pol & B_\pol \cr C_\pol & D_\pol}\end{equation}
where
\begin{eqnarray}& & A_{TE}=e^{i\beta_1a}\bkt{\cos\!\bkt{\beta_2b}+\frac{i}{2}\bkt{\frac{\beta_1}{\beta_2}+\frac{\beta_2}{\beta_1}}\sin\!\bkt{\beta_2b}}\\
& & B_{TE}=\frac{i}{2}e^{-i\beta_1a}\bkt{\frac{\beta_2}{\beta_1}-\frac{\beta_1}{\beta_2}}\sin\!\bkt{\beta_2b}\\
& & C_{TE}=\frac{i}{2}e^{i\beta_1a}\bkt{\frac{\beta_1}{\beta_2}-\frac{\beta_2}{\beta_1}}\sin\!\bkt{\beta_2b}\\
& & D_{TE}=e^{-i\beta_1a}\bkt{\cos\!\bkt{\beta_2b}-\frac{i}{2}\bkt{\frac{\beta_1}{\beta_2}+\frac{\beta_2}{\beta_1}}\sin\!\bkt{\beta_2b}}\end{eqnarray}
and
\begin{eqnarray}& & A_{TM}=e^{i\beta_1a}\bkt{\cos\!\bkt{\beta_2b}+\frac{i}{2}\bkt{\frac{n_2^2\beta_1}{n_1^2\beta_2}+\frac{n_1^2\beta_2}{n_2^2\beta_1}}\sin\!\bkt{\beta_2b}}\\
& & B_{TM}=\frac{i}{2}e^{-i\beta_1a}\bkt{\frac{n_1^2\beta_2}{n_2^2\beta_1}-\frac{n_2^2\beta_1}{n_1^2\beta_2}}\sin\!\bkt{\beta_2b}\\
& & C_{TM}=\frac{i}{2}e^{i\beta_1a}\bkt{\frac{n_2^2\beta_1}{n_1^2\beta_2}-\frac{n_1^2\beta_2}{n_2^2\beta_1}}\sin\!\bkt{\beta_2b}\\
& & D_{TM}=e^{-i\beta_1a}\bkt{\cos\!\bkt{\beta_2b}-\frac{i}{2}\bkt{\frac{n_2^2\beta_1}{n_1^2\beta_2}+\frac{n_1^2\beta_2}{n_2^2\beta_1}}\sin\!\bkt{\beta_2b}}\end{eqnarray}
As a result, \bref{Mgenright} is simplified to
\begin{equation}\smatrix{a_{N,\pol} \cr b_{N,\pol}}=M_\pol^{N}\smatrix{a_{0,\pol} \cr b_{0,\pol}}\label{aNbN=MNa0b0}\end{equation}
(for more elegance the index $R$ is omitted). Matrix $M_\pol$ is unimodular, therefore is obeys
the Chebyshev identity:
\begin{equation}M_\pol^N=\smatrix{A_\pol & B_\pol \cr C_\pol & D_\pol}^N=\smatrix{A_\pol U_{N,\pol}-U_{N-1,\pol} & B_\pol U_{N,\pol} \cr C_\pol U_{N,\pol} & D_\pol U_{N,\pol}-U_{N-1,\pol}}\label{Chebyshev}\end{equation}
with
\begin{eqnarray}U_{N,\pol}=\frac{\sin\!\bkt{N\eta_\pol}}{\sin\!\eta_\pol}& , & \eta_\pol=\arccos\!\frac{A_\pol+D_\pol}{2}\end{eqnarray}
what allows to simplify \bref{aNbN=MNa0b0} even more. Relation \bref{aNbN=MNa0b0} (or \bref{Mgenright}
and \bref{Mgenleft} in more general case) is a set of two linear equations\cite{inz}. Amplitudes $a_{N,\pol}$ and
$b_{N,\pol}$ are bound by external conditions, in this case it is just
\begin{equation}b_{N,\pol}=0\label{bN=0}\end{equation}
because we assume excitation in the photonic crystal (or at least on the other side of it) and there
is no reflection in the infinity. Because $a_{0,\pol}$ is the amplitude of the wave incident on layer boundary
in $z=a$, $b_{0,\pol}$ can be treated as reflected wave. Therefore, we can calculate reflection coefficient
of the whole part of photonic crystal reflecting $a_{0,\pol}$ wave, it will be
\begin{equation}r_\pol=\frac{b_{0,\pol}e^{-i\beta_1a}}{a_{0,\pol}e^{i\beta_1a}}\end{equation}
Knowing \bref{aNbN=MNa0b0}, \bref{Chebyshev} and \bref{bN=0} (and again marking index $R$ for ``right''
part of the crystal), we find that
\begin{equation}r_{R,\pol}=\frac{B_\pol}{D_\pol-\frac{\sin\!\bkt{\bkt{N_R-1}\eta_\pol}}{\sin\!\bkt{N_R\eta_\pol}}}\end{equation}
Because the structure is symmetrical, we instantly have expression for ``left'' part reflection coefficient
\begin{equation}r_{L,\pol}=\frac{B_\pol}{D_\pol-\frac{\sin\!\bkt{\bkt{N_L-1}\eta_\pol}}{\sin\!\bkt{N_L\eta_\pol}}}\end{equation}
(in general case of defected crystal it is not possible to obtain such a concise expression for
at least one reflection coefficient).

Having calculated reflection coefficients, we treat the layer of photonic crystal as a resonator,
with mirrors constituted by the remaining layers of the crystal and described by derived formulas.

\section{Mode spectrum and density of states in a layer of one-dimensional photonic crystal}
Let operator $\hat Q$ relate electromagnetic plane-wave $\vec E^{ex}$
exciting the resonator to effective plane-wave $\vec E^{eff}$ with the same wave
vector, that appears in the resonator due to excitement:
\begin{equation}\vec E^{eff}=\hat Q\vec E^{ex}\label{Qdef}\end{equation}
We call a plane-wave $\vec E_\pol\od{\vec k}$ with polarization $\pol$ and wave vector
$\vec k$ a mode of the resonator if it satisfies eigenequation of $\hat Q$:
\begin{equation}\hat Q\vec E_\pol\od{\vec k}=\Qe{\pol}{\vec k}\vec E_\pol\od{\vec k}\end{equation}
If the resonator is excited with plane-wave $\vec E_\pol\od{\vec k}$, effective
distribution of electromagnetic field is superposition of plane-wave $\Qe{\pol}{\vec k}\vec E_\pol\od{\vec k}$
with its coupled mode (i.e. plane wave with wave vector $\smatrix{k_x,k_y,-k_z}$), which appears because of reflections. Therefore, introduced
definition is equivalent to ``classical'' definition of a mode.

We introduce mode spectrum $\ms{\pol}{\vec k}$, related to distribution of modes in
wave vector domain. Thus, it can be used to calculate the number of modes $\NOS{k}$
with propagation constants less than $k$:
\begin{equation}\NOS{k}=\sum_\pol\lid{k'}{0}{k}k'^2\lid{\Omega'}{4\pi}{}\ms{\pol}{\vec k'}\label{NOS}\end{equation}
Differential of $\NOS{k}$ is known as density of states. From \bref{NOS} it can be
found, that
\begin{equation}\DOS{k}dk=k^2dk\sum_\pol\lid{\Omega}{4\pi}{}\ms{\pol}{\vec k}\label{DOS}\end{equation}
Because operator $\hat Q$ can be used to find distribution of modes, mode
spectrum $\ms{\pol}{\vec k}$ has to be proportional to its eigenvalues:
\begin{equation}\ms{\pol}{\vec k}=N_\mslett\Qe{\pol}{\vec k}\end{equation}
From \bref{DOS} we can calculate, that in free space, where $\Qe{\pol}{\vec k}=1$,
\begin{equation}\DOSfs{k}dk=k^2dk\sum_\pol\lid{\Omega}{4\pi}{}N_\mslett=8\pi N_\mslett k^2dk\label{DOSfs1}\end{equation}
It is a well known fact, that
\begin{equation}\DOSfs{k}dk=\frac{k^2}{\pi^2}dk\label{DOSfs2}\end{equation}
Comparing \bref{DOSfs1} and \bref{DOSfs2} we find the normalization constant $N_\mslett$:
\begin{equation}N_\mslett=\frac{1}{\bkt{2\pi}^3}\end{equation}
Finally, mode spectrum is defined as
\begin{equation}\ms{\pol}{\vec k}=\frac{\Qe{\pol}{\vec k}}{\bkt{2\pi}^3}\label{msdef}\end{equation}
Mode spectrum depicts influence of a FP cavity on a plane-wave. It also contains
information on anisotropy of the structure, and therefore is a magnitude more
fundamental than density of states.

\begin{figure}[!htpb]
\begin{center}
\includegraphics[width=.6\textwidth]{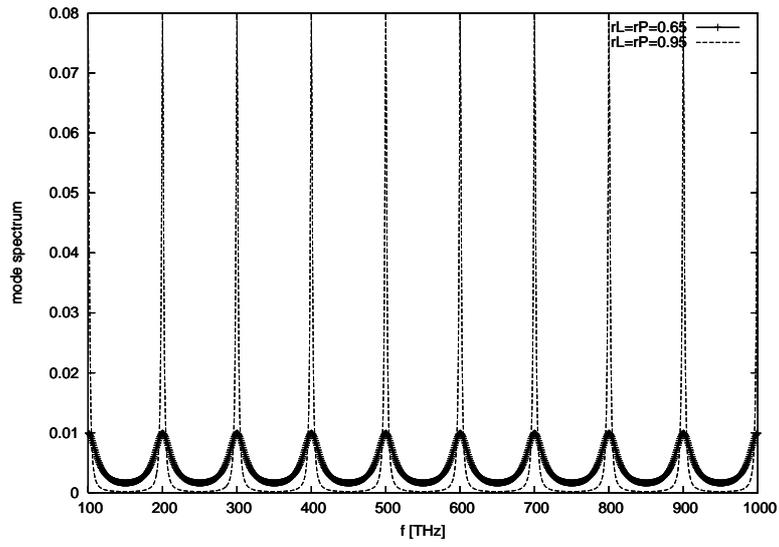}
\end{center}
\caption{Modes of FP resonator with constant reflection coefficients of mirrors.}
\label{peaks}
\end{figure}

To find mode spectrum of a $n_1$ layer of one-dimensional photonic crystal we need to find operator $\hat Q$
for a FP resonator filled with medium with refractive index $n_1$, built of two parallel mirrors at
distance $a$ from each other, with reflection coefficients $r_{R,\pol}$ and $r_{L,\pol}$. If the resonator is
excited with a plane-wave with wave vector $\vec k$ and amplitude of electric field $E_0$ (this
is a scalar problem), then amplitude of electric field of effective plane-wave
with wave vector $\vec k$ can be found by summation of two infinite series\cite{mgr,inz}:
\begin{equation}E^{eff}=E_0+\sum_{j=1}^\infty r_{R,\pol}r_{L,\pol}e^{2i\beta a}E_0+\sum_{j=1}^\infty r_{R,\pol}^*r_{L,\pol}^*e^{-2i\beta a}E_0\end{equation}
where $\beta=\beta_1$ defined by \bref{beta} and $^*$ means complex conjugation.
Comparing to definition \bref{Qdef} we can conclude, that the operator $\hat Q$ for the resonator is:
\begin{equation}\hat Q=\frac{1-\abs{r_{R,\pol}r_{L,\pol}}^2}{1+\abs{r_{R,\pol}r_{L,\pol}}^2-2\real{r_{R,\pol}r_{L,\pol}e^{2i\beta a}}}\end{equation}
and from \bref{msdef} we obtain the formula for mode spectrum of the photonic crystal's layer:
\begin{equation}\ms{\pol}{\vec k}=\frac{1}{8\pi^3}\frac{1-\abs{r_{R,\pol}r_{L,\pol}}^2}{1+\abs{r_{R,\pol}r_{L,\pol}}^2-2\real{r_{R,\pol}r_{L,\pol}e^{2i\beta a}}}\end{equation}
To find density of states we can use definition \bref{DOS}. In spherical coordinates $\beta=k\cos\!\vartheta$, therefore
\begin{equation}\DOS{k}=k^2\lid{\vartheta'}{0}{\pi}\sin\!\vartheta'\lid{\varphi'}{0}{2\pi}\sum_{\pol}\frac{1}{8\pi^3}\frac{1-\abs{r_{R,\pol}r_{L,\pol}}^2}{1+\abs{r_{R,\pol}r_{L,\pol}}^2-2\real{r_{R,\pol}r_{L,\pol}e^{2iak\cos\!\vartheta'}}}\end{equation}
Integration over $\varphi'$ can be done, because of rotational symmetry of the crystal. Additionally
substituting \mbox{$u=\cos\!\vartheta'$} we reach the following formula:
\begin{equation}\DOS{k}=\frac{k^2}{4\pi^2}\lid{u}{-1}{1}\sum_{\pol}\frac{1-\abs{r_{R,\pol}r_{L,\pol}}^2}{1+\abs{r_{R,\pol}r_{L,\pol}}^2-2\real{r_{R,\pol}r_{L,\pol}e^{2iaku}}}\end{equation}
The above expression is true for any resonator with two parallel mirrors. In particular, taking
$r_{R,\pol}=r_{L,\pol}=0$ and $n_1=1$ we obtain expression for density of states in free space:
\begin{equation}\DOSsign_{fs}\!\bkt{k}=\frac{k^2}{\pi^2}dk\end{equation}
If both $r_{L,\pol}$ and $r_{R,\pol}$ are constant, mode spectrum takes form
of regularly distributed ``peaks'' (see figure \ref{peaks}).

\section{Variation of mode spectrum in a one-dimensional photonic crystal}
\begin{figure}[!htpb]
\begin{center}
\includegraphics[width=.6\textwidth]{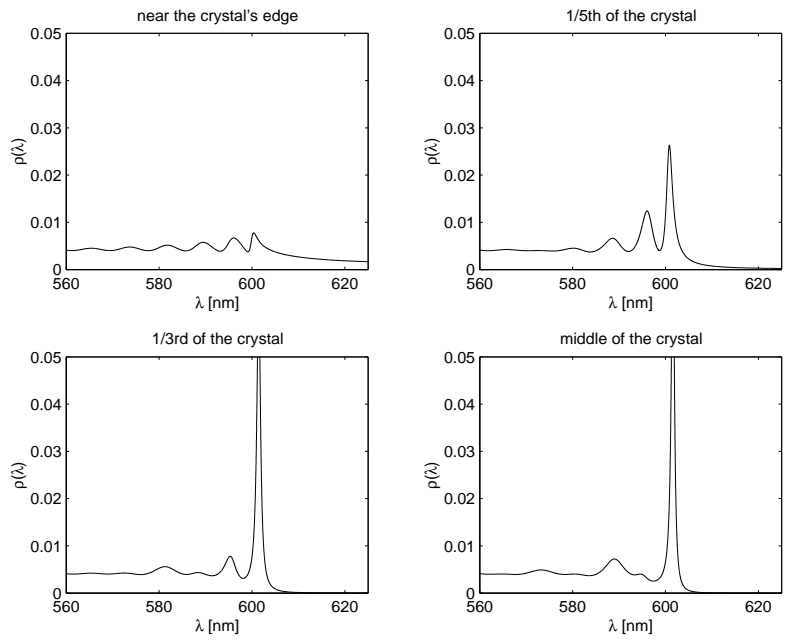}
\end{center}
\caption{Edge of photonic band gap in different layers of a 1D photonic crystal.}
\label{phbedge}
\end{figure}

\begin{figure}[!htpb]
\begin{center}
\includegraphics[width=.68\textwidth]{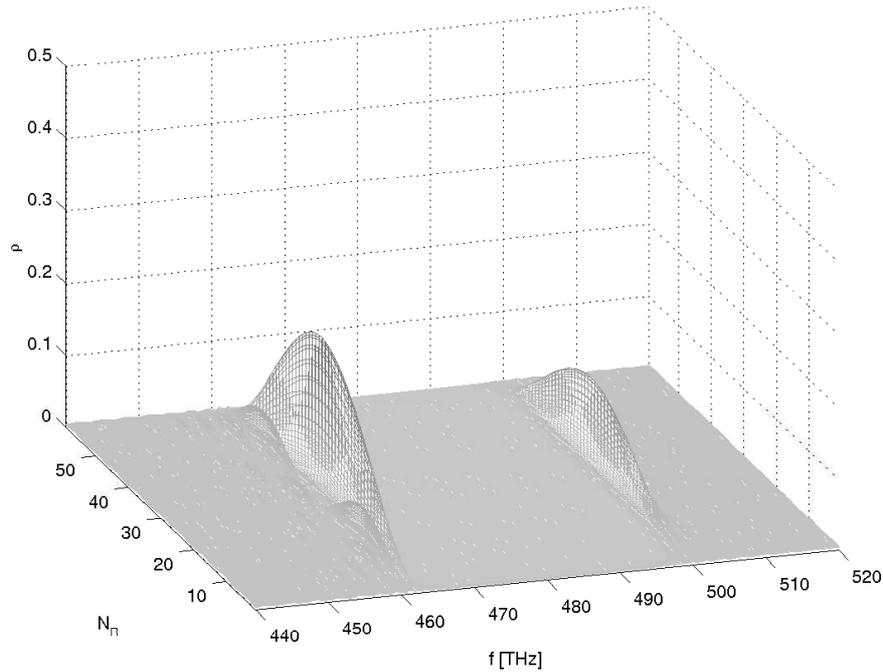}
\end{center}
\caption{Mode spectrum in a 1D photonic crystal.}
\label{ms1d}
\end{figure}

Presented method of mode spectrum calculation allows to analyze its space dependence in photonic crystal
and therefore also the crystals properties. For example, it is possible to examine the photonic
band gap edge in different layers of the crystal (figure \ref{phbedge}). It can be clearly seen, that
in layers (sufficiently) close to the end of the photonic crystal full photonic band gap does not
form. The reason of this behavior is that reflection coefficients $r_L$ and $r_P$
depend on number of layers surrounding selected layer. If this number is high,
amplitude of reflection coefficient can be close to $1$, while for low number of
layers amplitude of reflection coefficient is significantly lower. In the second
case, quality of resonator is low and its mode spectrum tends to be more similar
to mode spectrum of free space. Therefore, formation of distinctive photonic
band gap does not occur. This regularity can be as well observed on figure
\ref{ms1d}, which depicts mode spectrum (for perpendicularly incident
plane-wave) in different layers of the crystal. Amplitude of mode spectrum
at the edge of photonic band gap reaches highest values in the middle of
the crystal, where amplitudes of reflection coefficients $r_L$ and $r_P$,
along with effective resonator quality, have the highest values. Another
interesting observation, is that photonic band gap in general is not symmetrical
-- amplitudes of mode spectrum at the edges of the gap are not equal.

\bibliography{SPIE_ref}
\acknowledgments
Project granted by the Ministry of Science and Information Society Technologies
in Poland, project no 3 T11B 069 28 and the Sixth European Union Framework
Programme for Research and Technological Development (FP6) -- Network of Excellence
on Micro-Optics ``NEMO''.

\end{document}